\def\OPTIONLoudLabels{0}

\newcommand{\ifproofs}{\iffalse}

\documentclass[publicdomain]{eptcs}  %
\newcommand{\event}{2010 Workshop on Intersection Types and Related Systems (ITRS '10)}

\usepackage{relsize}

\newcommand{\keywordfontsize}{9pt}  %
\newcommand{\keywordfontstep}{-0}

\usepackage[small,bf]{caption2}
\usepackage{color}
\usepackage{xcolor}
\usepackage{pstricks,pst-node,graphicx,eepic}
\usepackage{url}
\usepackage{breakurl}
\usepackage{listings}
\usepackage{refcount}

\usepackage{natbib} \bibpunct{(}{)}{;}{a}{}{,}

\usepackage{xspace}
\renewcommand{\ttdefault}{cmtt}   %
\renewcommand{\sfdefault}{cmss}   %

\usepackage{pifont}
\usepackage{url}
\usepackage{enumerate}
\usepackage{comment}
\usepackage{proof}
\usepackage{framed}
\usepackage{rotating}
\DeclareMathVersion{bold}
\usepackage{amssymb}
\usepackage{amsmath}
\usepackage{amsthm}

\usepackage{mathptmx,bm}
\let\MathRightArrow\Rightarrow %
\usepackage{marvosym} %
\def\Rightarrow{\MathRightArrow}

\newtheoremstyle{nonstupid}%
   {\topsep}%
   {\topsep}%
   {\upshape\slshape}%
   {}%
   {\bfseries}%
   {.}%
   {0.7em}%
   {}%
\theoremstyle{nonstupid}
\newtheorem{theorem}{Theorem}

\newtheorem{lemma}[theorem]{Lemma}
\newtheorem{corollary}[theorem]{Corollary}
\newtheorem*{corollary*}{Corollary}

\newtheorem*{conjecture*}{Conjecture}
\newtheorem{exercise}[theorem]{Exercise}
\newtheorem{example}[theorem]{Example}
\newtheorem*{example*}{Example}

\newtheorem*{theorem*}{Theorem}
\theoremstyle{remark} 
\theoremstyle{definition} \newtheorem{definition}[theorem]{Definition}

\usepackage{srcltx}

\newcommand{\xLam}{\lambda}
\newcommand{\Lam}[1]{\xLam#1.\,}
\newcommand{\xFix}{\ensuremath{\keyword{fix}}}
\newcommand{\Fix}[1]{\xFix~#1.\:}
\newcommand{\arr}{\rightarrow}
\newcommand{\entails}{\,\vdash\,}         %

\newcommand{\subtype}{\mathrel{\leq}}       %
\newcommand{\sectty}{\mathrel{\land}}          %
\newcommand{\unty}{\mathrel{\lor}}          %
\newcommand{\topty}{\top}
\newcommand{\botty}{\bot}
\newcommand{\UnivSym}{\Pi}
\newcommand{\ExisSym}{\Sigma}

\newcommand{\step}{\mapsto}

\newcommand{\xCase}{\ensuremath{\keyword{case}}}
\newcommand{\Case}[2]{{\xCase~#1~\keyword{of}~#2}}

\newcommand{\Let}[2]{{\xxLet\:#1\,{\texttt{=}}\,#2\:\keyword{in}\:}}

\newcommand{\textfn}[1]{\textsf{#1}}

\newcommand{\tyname}[1]{\mathsf{#1}}
\newdimen\zzkwfontsize
\newcommand{\keyword}[1]{\zzkwfontsize=\keywordfontsize%
\text{\usefont{T1}{\sfdefault}{r}{n}%
\relsize{\keywordfontstep}\selectfont#1}}
\newcommand{\textkw}[1]{\keyword{#1}}

\newcommand{\Int}{\tyname{int}}
\newcommand{\Bool}{\tyname{bool}}

\newcommand{\SomeInt}{\tyname{some}}
\newcommand{\None}{\tyname{none}}

\newcommand{\arrayenvb}[1]{\renewcommand{\arraystretch}{1}  \begin{array}[b]{@{}c@{}}#1\end{array}}

\newcommand{\Tbool}{\Bool}

\newcommand{\Unit}{\tyname{unit}}

\newcommand{\against}{\Downarrow}
\newcommand{\has}{\Uparrow}

\newcommand{\Dee}{\mathcal{D}}

\newcounter{codeLineCntr}

\newcommand{\out}[1]{}

\renewcommand{\phi}{\varphi}

\newcommand{\Figureref}[1]{Figure \ref{#1}}

\newcommand{\Sectionref}[1]{Section \ref{#1}}

\newcommand{\Theoremref}[1]{Theorem \ref{#1}}

\newcommand{\Lemmaref}[1]{Lemma \ref{#1}}

\newcommand{\Propositionref}[1]{Proposition \ref{#1}}

\newcommand{\barvar}[1]{\overline{\mathbf{#1}}}
\newcommand{\xbarvar}{\barvar{x}}
\newcommand{\ybarvar}{\barvar{y}}

\newcommand{\val}{~\mathsf{value}}
\newcommand{\AND}{\textrm{~and~}}

\newcommand{\hole}{\ensuremath{[\,]}}

\newcommand{\etal}{{et al.}}
\newcommand{\eg}{e.g.\ }

\newcommand{\Lbrack}{\char"5B}
\newcommand{\Rbrack}{\char"5D}

\definecolor{lred}{rgb}{1.0, 0.3, 0.3}
\newrgbcolor{lred}{1.0 0.3 0.3}

\newcommand{\BLACKNODE}[1]{~~\raise4pt\hbox{\psellipse[fillstyle=solid,fillcolor=black](0, 0)(6pt,6pt)}\hspace{-3pt}{\textcolor{white}{\textbf{#1}}}~\:}
\newcommand{\REDNODE}[1]{~~\raise4pt\hbox{\psellipse[fillstyle=solid,fillcolor=lred](0, 0)(6pt,6pt)}\hspace{-3pt}{\textbf{#1}}~\:}

\newlength\zzskipwidthlen

\newcommand{\textt}[1]{\texttt{1}}

\newdimen{\zzzpbox}

\newdimen\zzfontsz
\newcommand{\fontsz}[2]{\zzfontsz=#1%
{\fontsize{\zzfontsz}{1.2\zzfontsz}\selectfont{#2}}}

\newcommand{\mathsz}[2]{\text{\fontsz{#1}{$#2$}}}

\newcommand{\rampfont}[2]{\zzfontsz=\f@encoding plus #1%
{\fontsize{\zzfontsz}{1.2\zzfontsz}\selectfont{#2}}}

\newlength{\zzsplatboldwidth}
\newcommand{\xsplatbold}[2]{\settowidth{\zzsplatboldwidth}{{#2}}{#2}\addtolength{\zzsplatboldwidth}{-#1}\hspace{-\zzsplatboldwidth}\raisebox{#1}{{#2}}}
\newcommand{\splatbold}[1]{\xsplatbold{-0.04mm}{#1}}

\makeatletter
\def\url@MGstyle{%
\def\UrlFont{\tiny\ttfamily}%
\def\do@url@hyp{\do\-}%
\Url@do
}
\makeatother

\ifnum\OPTIONLoudLabels=1%
\newcommand{\Label}[1]{\LoudLabel{#1}}%
\else%
\newcommand{\Label}[1]{\label{#1}}%
\fi

\newcommand{\Infer}[3]{\infer[\text{\strut#1}]{#3\mathstrut}{#2\mathstrut}}

\newcommand{\InferAnon}[2]{\infer{#2}{#1\mathstrut}}

\newcommand{\BeginProof}{\renewcommand{\arraystretch}{1.1} \begin{tabular}[b]{r@{}r @{} l @{~~~} l}}
\newcommand{\EndProof}{\end{tabular} \renewcommand{\arraystretch}{\mydefaultarraystretch}}

\newenvironment{llproof}{\BeginProof}{\EndProof}

\newcommand{\proofheading}[1]{}  %

\newcommand{\ctxsubtype}{\mathrel{\,\lesssim\,}}

\newcommand{\Rctxanno}{\ensuremath{\mathsf{ctx}\textsf{-}\mathsf{anno}}\xspace}

\newcommand{\againstL}{\mathrel{\against^{\mathbb{L}}}}

\newcommand{\Rlet}{\ensuremath{\mathsf{let}}\xspace}

\newcommand{\Rdirectleft}{\ensuremath{\mathsf{direct} \mathbb{L}}\xspace}
\newcommand{\Rleftbot}{\ensuremath{\bot \mathbb{L}}\xspace}

\newcommand{\Rleftun}{\ensuremath{\lor \mathbb{L}}\xspace}

\newcommand{\Rleftsect}[1]{\ensuremath{\land \mathbb{L}_{#1}}\xspace}

\newcommand{\X}{\xbarvar}
\newcommand{\Y}{\ybarvar}
\newcommand{\Z}{\barvar{z}}
\newcommand{\xxLet}{\ensuremath{\mathsf{let}}\xspace}

\newcommand{\skull}{\ensuremath{\text{\large$\sim$}}}

\newcommand{\littleskull}{\ensuremath{\text{\footnotesize$\skull$}}}

\newcommand{\Rslackvar}{\ensuremath{\littleskull\mathsf{var}}\xspace}

\newcommand{\Rletslack}{\ensuremath{\mathsf{let}{\littleskull}}\xspace}
\newcommand{\Letslack}[2]{\ensuremath{\keyword{let}\:{\littleskull}#1\,\texttt{=}\,#2\:\keyword{in}\:}}
\newcommand{\slackbind}[2]{\ensuremath{\littleskull#1}{\,\texttt{=}\,}{#2}}

\newcommand{\linear}{\Vdash}

\newcommand{\gokentails}{\entails} %
\newcommand{\cok}{\ensuremath{\mathsf{~ok}}}
\newcommand{\gok}{\cok}  %

\newgray{grayboxgray}{0.85}
\newcommand{\textgraybox}[1]{\psframebox[framesep=0pt,fillcolor=grayboxgray,linewidth=0.5pt]{\fbox{\parbox[s][\totalheight]{0mm}{}#1}}}

\newcommand{\textshadebox}[1]{\psframebox[framesep=0pt,fillstyle=solid,fillcolor=grayboxgray,linestyle=none,linecolor=white]{\parbox[s][\totalheight]{0mm}{}#1}}
\newcommand{\shadebox}[1]{\textshadebox{\ensuremath{#1}}}

\newcommand{\judgbox}[1]{\textgraybox{\large\ensuremath{#1}}}

\newcommand{\lettrans}{\mathrel{\,\hookrightarrow\,}}

\newcommand{\freein}{\ensuremath{~\keyword{in}~}}   %
\newcommand{\bind}[2]{#1{\,\texttt{=}\,}#2}

\newcommand{\unwindingx}{{\hookleftarrow}} %

\newcommand{\unw}[1]{\unwindingx_{\!{}}(#1)}
\newcommand{\E}{\mathcal{E}}
\newcommand{\Elong}{\mathcal{Q}}

\newcommand{\Rapp}{\ensuremath{{\arr}\text{E}}\xspace}
\newcommand{\Rlam}{\ensuremath{{\arr}\text{I}}\xspace}

\newcommand{\Rfix}{\ensuremath{\mathsf{fix}}\xspace}

\newcommand{\Rsectintro}{\ensuremath{\land \text{I}}\xspace}
\newcommand{\Rsectelim}[1]{\ensuremath{\land \text{E}_{#1}}\xspace}

\newcommand{\Runintro}[1]{\ensuremath{\lor \text{I}_{#1}}\xspace}
\newcommand{\Runelim}{\ensuremath{\lor \text{E}}\xspace}

\newcommand{\Rvar}{\ensuremath{\mathsf{var}}\xspace}
\newcommand{\Rfixvar}{\ensuremath{\mathsf{fixvar}}\xspace}
\newcommand{\Rsub}{\ensuremath{\mathsf{sub}}\xspace}

\newcommand{\Rvarbar}{\ensuremath{\overline{\mathsf{var}}}\xspace}

\newcommand{\MKSUB}[1]{\ensuremath{{#1}{\subtype}}\xspace}
\newcommand{\subArr}{\MKSUB{\arr}}

\newcommand{\subSectL}[1]{\MKSUB{{\sectty}\text{L}\ensuremath{_{#1}}}}
\newcommand{\subSectR}{\MKSUB{{\land}\text{R}}}

\newcommand{\subUnionL}{\MKSUB{{\lor}\text{L}}}
\newcommand{\subUnionR}[1]{\MKSUB{{\unty}\text{R}\ensuremath{_{#1}}}}
\newcommand{\subBotL}{\MKSUB{{\botty}\text{L}}}

\newcommand{\cmtbegin}{\texttt{(*}}
\newcommand{\cmtend}{\texttt{*)}}
\newcommand{\annobegin}{\cmtbegin\texttt{\Lbrack}~~\,}
\newcommand{\annoend}{\,~~\texttt{\Rbrack}\cmtend}

\newcommand{\xitrans}[1]{{|}#1{|}}

\newcommand{\oneinferencerule}[1]{

  \medskip

  \centerline{#1}
  \medskip
  \noindent\hspace{-0.047in}}

\newcommand{\mydefaultarraystretch}{1.2}

\newcommand{\AntiV}{\widehat{e}}
\newcommand{\PreV}{\check{e}}

\newcommand{\caret}{\char94}

\newenvironment{displ}{\vspace{2pt} \begin{center}\renewenvironment{tabular}{\begin{tabular}}{\end{tabular}} ~\!\!}{\end{center}\vspace{2pt}}
\newcommand{\vertcenter}[1]{\begin{tabular}{c}#1\end{tabular}}

\newcommand{\marginnoteRmlExamples}[1]{{}}

\newcounter{zzInOinkComment}
\newcommand{\oinkkw}[1]{\ifnum\value{zzInOinkComment}=0{\usefont{T1}{cmtt}{m}{it}{\splatbold{#1}}}\else{\normalfont\textsl{#1}}\fi}

\newcommand{\lladdconj}{\mathrel{\binampersand}}

\newcommand{\RZZaddconjR}[1]{\ensuremath{{\lladdconj}\text{R}}\xspace}

\newdimen\zzlistingsize
\newdimen\zzlistingsizedefault
\zzlistingsize=\zzlistingsizedefault
\global\def\CommentCopter{0}
\newcommand{\Lstbasicstyle}{\fontsize{\zzlistingsize}{1.05\zzlistingsize}\ttfamily%
}
\newcommand{\keywordcopter}{\fontsize{0.95\zzlistingsize}{1.0\zzlistingsize}\bf}
\newcommand{\stupidcopter}{\if0\CommentCopter\keywordcopter\fi}
\newcommand{\commentcopter}{\def\CommentCopter{1}\fontsize{0.95\zzlistingsize}{1.0\zzlistingsize}\rmfamily\slshape}

\newlength{\zzlstwidth}
\newcommand{\setlistingsize}[1]{\zzlistingsize=#1%
\settowidth{\zzlstwidth}{{\Lstbasicstyle~}}%
}
\setlistingsize{\zzlistingsizedefault}

\renewcommand{\Phi}{\text{\fontsz{40pt}{PHI}}}

\definecolor{dRed}{rgb}{0.65, 0.0, 0.0}
\definecolor{DRED}{rgb}{0.65, 0.0, 0.0}
\definecolor{dGreen}{rgb}{0.0, 0.65, 0.0}
\definecolor{dBlue}{rgb}{0.0, 0.0, 0.65}
\definecolor{dPurple}{rgb}{0.65, 0.0, 0.65}
\definecolor{dDigPurple}{rgb}{0.5, 0.0, 0.5}
\definecolor{DDIGPURPLE}{rgb}{0.5, 0.0, 0.5}  %
\definecolor{dFaint}{rgb}{0.7, 0.7, 0.7}
\definecolor{dGray}{rgb}{0.5, 0.5, 0.5}
\definecolor{dDark}{rgb}{0.2, 0.2, 0.2}
\definecolor{dAlmostBlack}{rgb}{0.1, 0.1, 0.1}
\definecolor{dExColor}{rgb}{0.3, 0, 0.3}
\definecolor{DEXCOLOR}{rgb}{0.3, 0.0, 0.3}  %
\definecolor{dExColorIn}{rgb}{0, 0, 0.1}
\definecolor{dExColorOut}{rgb}{0, 0, 0.1}

\newcommand{\XBeginProof}{\vspace{5pt} \renewcommand{\arraystretch}{1.1} \begin{tabular}[b]{r@{}r @{}c@{}l@{~}l l }}

\title{Untangling Typechecking of Intersections and Unions}
\newcommand{\Counter}[2]{\relax}

\newcommand{\titlerunning}{Untangling Typechecking of Intersections and Unions}
\newcommand{\authorrunning}{J.\ Dunfield}

\author{Joshua Dunfield\footnote{Current address:
MPI-SWS, Gottlieb-Daimler-Str.\ 49, 67663 Kaiserslautern, Germany.}
\institute{School of Computer Science, McGill University \\ Montr\'eal, Canada}
\email{joshua@mpi-sws.org}%
}

\begin{document}
\maketitle

\begin{abstract}
  Intersection and union types denote conjunctions and disjunctions of properties.
  Using bidirectional typechecking, intersection types are relatively
  straightforward, but union types present challenges.  For union types, we can case-analyze
  a subterm of union type when it appears in evaluation position (replacing the subterm
  with a variable, and checking that term twice under appropriate assumptions).
  This technique preserves soundness in a call-by-value semantics.
  Sadly, there are so many choices of subterms that a direct implementation is not
  practical.  But carefully transforming programs into let-normal form drastically
  reduces the number of choices.  The key results are soundness and
  completeness: a typing derivation (in the system with too many subterm choices) exists
  for a program if and only if a derivation exists for the let-normalized program.
\end{abstract}

\setcounter{footnote}{0}

\section{Introduction}

To check programs in advanced type systems, it can be useful to split the traditional
typing judgment $e : A$ into two forms, $e \has A$ read ``$e$ synthesizes type $A$'' and $e
\against A$ read ``$e$ checks against type $A$'', and requiring that the user write annotations on
redexes.  This \emph{bidirectional typechecking}~\citep{Pierce98popl} is decidable for many interesting
features, including intersection and union types without syntactic markers.
\emph{Tridirectional} typechecking~\citep{Dunfield04:Tridirectional,DunfieldThesis}
is essentially bidirectional, but union types are eliminated with the aid of
a \emph{tridirectional rule} that uses an evaluation
context $\E$:
\begin{displ}
  \Infer{\Rdirectleft}
    {\Gamma; \Delta_1 \entails e' \has A
    ~~&~~  \Gamma; \Delta_2, \X{:}A \entails \E[\X] \against C}
    {\Gamma; \Delta_1, \Delta_2 \entails \E[e'] \against C}
\end{displ}
In this rule, $\Gamma$ is an ordinary variable context and $\Delta_1, \Delta_2$ is the concatenation of
\emph{linear contexts}; linear variables $\X$ in $\Delta$s essentially stand for subterms (occurrences) in
the subject.  \Rdirectleft gives $e'$ a (linear) name $\X$, so that a left rule, which decomposes types in
the context $\Delta$, can eliminate union types appearing in $A$.  Instead of a direct union elimination rule like \Runelim, we use \Rdirectleft together with a left rule \Rleftun.
\begin{displ}
 \small
          \Infer{\!\text{[\Runelim{}]}}
                  {\Gamma \entails e' \has A \unty B
               ~&~ \Gamma, x{:}A \entails \E[x] \against C
               ~& \Gamma, y{:}B \entails \E[y] \against C}
                  {\Gamma \entails \E[e'] \against C}
    ~~~
     \Infer {\!\Rleftun}
          {\Gamma; \Delta, \X{:}A \entails e \against C
            ~&~ \Gamma; \Delta, \X{:}B \entails e \against C}
          {\Gamma; \Delta, \X{:}A \unty B \entails e \against C}  
\end{displ}

While evaluation contexts are defined syntactically, this rule is not
syntax-directed in the usual sense: many terms have more than one
decomposition into some $\E[e']$ where the subterm %
$e'$ can synthesize a type.  Under a left-to-right (functions first, arguments second) call-by-value semantics,
even $f~x$ has three decompositions $\E = \hole$, $\E = \hole~x$, $\E = f~\hole$,
so a straightforward implementation of a system with \Rdirectleft
would require far too much backtracking.  Compounded with
backtracking due to intersection and union types (e.g., if $f : (A_1 \arr A_2)
\sectty (B_1 \arr B_2)$ we may have to try both $f : A_1 \arr A_2$ and
$f : B_1 \arr B_2$), such an implementation would be hopelessly impractical.

This paper reformulates tridirectional typechecking (summarized in
\Sectionref{sec:tridirectional}) to work on terms in a particular \emph{let-normal form},
in which steps of computation are sequenced and intermediate computations are named.
The let-normal transformation (\Sectionref{sec:let}) drastically constrains the
decomposition by sequencing terms, forcing typechecking to proceed left to right
(with an interesting exception).  The results stated in
Section~\ref{sec:results} guarantee
that the let-normal version of a program $e$ is well typed under the let-normal
version of the type system if and only if $e$ is well typed under the tridirectional
system.  %

The let-normal transformation itself is not complicated, though the motivation for my
particular formulation is somewhat involved.  The details of the transformation may be of
interest to designers of advanced type systems, whether their need for a sequential form
arises from typechecking itself (as in this case) or from issues related to
compilation.

Unfortunately, the proofs (especially the proof of completeness) are very involved; I
couldn't even fit all the statements of lemmas in this paper, much less sketch their proofs.
I hope only to convey a shadow of the argument's structure. %

This paper distills part of my dissertation~\citep[chapter 5]{DunfieldThesis}.
To simplify presentation, I omit tuples, datasort refinements,
indexed types (along with universal and existential quantification, guarded types, and asserting
types), and a greatest type $\topty$.

\section{Tridirectional Typechecking}  \Label{sec:tridirectional}

We have functions, products, intersections, unions, and an empty type $\botty$.  We'll use a $\Unit$ type
and other base types like $\Int$ in examples.  In the terms $e$, we have
variables $x$ (which are values) bound by $\Lam{x} e$, variables $u$ (not values)
bound by $\Fix{u} e$, and call-by-value application $e_1\,e_2$.  %
Note the lack of syntactic markers for intersections or unions.  %
As usual, $\E[e']$ is the evaluation context $\E$ with its hole replaced by $e'$.
To replace $x$ with $e_1$, we write $[e_1/x]e_2$: ``$e_1$ for $x$ in $e_2$''.
\begin{displ}
        \begin{array}[t]{r@{~}l}
             \text{Types}  ~~           A,B,C,D ~&~::=~~
                 A \arr B %
                            ~|~ A \sectty B
                            ~|~ A \unty B  ~|~ \botty
        \\[2pt]
             \text{Terms}  ~~    e &~::=~~ x ~|~ u
                       ~|~ \Lam{x}e ~|~ e_1\,e_2
                       ~|~ \Fix{u}e
        \\[2pt]
        \textrm{Values}~~~v&~::=~~ x
                   ~|~ \Lam{x}e    %
        \\[2pt]
        \textrm{Evaluation contexts}~~\, \E &~::=~~ \hole ~|~ \E\,e ~|~ v\;\E %
    \end{array}
    ~
    \begin{array}[t]{l} ~\\[-10pt] \framebox{\begin{array}[t]{l}
       \textrm{Small-step reduction rules} \\[2pt]
           \multicolumn{1}{l}{
              \begin{array}[t]{@{}r@{~~}c@{~~}l}
                \E[(\Lam{x}{e})\,v] &\step&  \E[\,[v/x]e\,]
              \\[3pt]
                 \E[\Fix{u}e] &\step&  \E[[(\Fix{u}e)\:/\:u]\,e] 
              \end{array}
           }
       \end{array}} \end{array}
\end{displ}

We'll start by looking at the ``left tridirectional'' (in this paper, called just ``tridirectional'') system.
This system was presented in \citet{Dunfield04:Tridirectional}
and \citet[chapter 4]{DunfieldThesis}; space allows only a cursory description.  %

The subtyping judgment (\Figureref{fig:subtyping}) is $A \subtype B$.  Transitivity is
admissible. $\sectty$ does not distribute across $\arr$, for reasons explained by
\citet{Davies00icfpIntersectionEffects}.

\Figureref{fig:tri-system} gives the typing rules.  The judgment $\Gamma; \Delta \entails
e \has A$ is read ``$e$ synthesizes type $A$'', and $\Gamma; \Delta \entails e \against A$ is read
``$e$ checks against $A$''.  When synthesizing, $A$ is output; when checking,
$A$ is input.

\begin{displ}
  \begin{array}[t]{r@{~}l}
         \text{Contexts}  ~~           \Gamma ~&~::=~~ \cdot ~|~ \Gamma, x{:}A
\\[2pt]     \text{Linear contexts}  ~~      \Delta ~&~::=~~  \cdot ~|~ \Delta, \X{:}A
   \end{array}
\end{displ}

Contexts $\Gamma$ have regular variable declarations.  Linear contexts $\Delta ::=
\cdot ~|~ \Delta, \X{:}A$ have linear variables.  If $\Gamma; \Delta \entails e \dots$ is
derivable, then ``$\Delta \linear e \gok$'', read ``$e$ OK under $\Delta$'':
each $\X$ declared in $\Delta$ appears
exactly once in $e$, and $e$ contains no other linear variables.  Rules that decompose the
subject, such as \Rapp decomposing $e_1\,e_2$ into $e_1$ and $e_2$, likewise decompose
$\Delta$.

Most rules follow a formula devised in \citet{Dunfield04:Tridirectional}: introduction
rules, such as \Rlam, check; elimination rules, such as \Rapp, synthesize.  Introduction
forms like $\Lam{x}e$ %
thus construct \emph{synthesizing terms},
while elimination forms like $e_1\,e_2$ %
are \emph{checked terms}. Some rules fall outside this classification.  The
\emph{assumption} rules \Rvar, \Rvarbar and \Rfixvar synthesize (an assumption
$x{:}A$ can be read $x{\has}A$).  The \emph{subsumption} rule \Rsub allows a term
that synthesizes $A$ to check against a type $B$, provided $A$ is a subtype of $B$.
The rule \Rctxanno permits \emph{contextual typing annotations}; for example, in
$(\textfn{succ}~x : (x{:}\textsf{odd} \entails \textsf{even}, x{:}\textsf{even}
\entails \textsf{odd}))$, the annotated term $\textfn{succ}~x$ is checked against
$\textsf{even}$ if $x{:}\textsf{odd} \in \Gamma$, and against $\textsf{odd}$ if
$x{:}\textsf{even} \in \Gamma$.  The premise $(\Gamma_0 \entails A) \ctxsubtype
(\Gamma \entails A)$ is derivable if the assumptions in $\Gamma$ support the
assumptions in $\Gamma_0$.  For details, see \citet{Dunfield04:Tridirectional}.

Finally, we have \emph{left rules}
\Rleftsect{1}, \Rleftsect{2}, \Rleftun, \Rleftbot which
act on linear assumptions $\X{:}A$ where $A$ is of intersection, union, or empty type.
These act as elimination rules---for $\unty$ and $\botty$, they are the \emph{only} elimination rules.
\Rleftsect{1} and \Rleftsect{2} are not useful alone (the ordinary eliminations \Rsectelim{1} and
\Rsectelim{2} would do) but are needed to expose a nested $\unty$ for \Rleftun, or a $\botty$ for \Rleftbot.

The backtracking required to choose between \Rsectelim{1} and \Rsectelim{2}, or between
\Runintro{1} and \Runintro{2}, or between the related subtyping rules, as well as the need
to check a single term more than once (\Rsectintro, \Rleftun) suggests that typechecking
is exponential.  In fact, \citet[pp.\ 67--68]{Reynolds96:Forsythe} proved that for a closely related
system, typechecking is \textsc{pspace}-hard.  We can't make typechecking polynomial,
but ``untangling'' \Rdirectleft will remove one additional source of complexity.

    \begin{figure}[htbp]
      \begin{center}
      \fontsz{10.75pt}{\begin{tabular}{l}
              \begin{tabular}[b]{l}\hspace{-12pt}\judgbox{A \subtype B} \\[-4pt] ~\end{tabular}
              \Infer {\subArr}
                  {B_1 \subtype A_1   ~&~~   A_2 \subtype B_2}
                  {A_1{\arr}A_2 \subtype B_1{\arr}B_2}
              ~~
              \Infer {\!\subSectL{1}}
                   {A_1 \subtype B}
                   {A_1{\sectty\,}A_2 \subtype B}
              ~
              \Infer {\!\subSectL{2}}
                   {A_2 \subtype B}
                   {A_1{\sectty\,}A_2 \subtype B}
              ~~
              \Infer {\!\subSectR}
                   {A \subtype B_1  ~~&~~ A \subtype B_2}
                   {A \subtype B_1{\sectty\,}B_2}
         \\[0.8em]~~~~~~~~~~~~~
             \Infer{\!\subBotL}
                  {}
                  {\botty \subtype A}
              ~~~~~
                \Infer{\!\!\subUnionL}
                    {A_1 \subtype B ~&~~ A_2 \subtype B}
                    {A_1{\unty}A_2 \subtype B}
                ~~~~~
                \Infer{\!\!\subUnionR{1}}
                    {A \subtype B_1}
                    {A \subtype B_1{\unty}B_2}
                ~~~~
                \Infer{\!\!\subUnionR{2}}
                    {A \subtype B_2}
                    {A \subtype B_1{\unty}B_2}
     \end{tabular}}
     \end{center}
~\\[-2.7em] $~$
    \caption{Subtyping}
    \label{fig:subtyping}
    \index{subtyping!summary of rules}
\end{figure}

   \begin{figure}[th]
 ~\\[-18pt]
   \centering

     \begin{tabular}{@{}c@{}}
       \fontsz{10pt}{
        \begin{tabular}{@{}c@{}}
          \Infer {\Rvar}
            {\Gamma(x) = A}
            {\Gamma; \cdot \entails x \has A}
           ~~~
            \Infer{\Rvarbar}
            {}
            {\Gamma; \X{:}A \entails \X \has A}
          ~~~~
          \Infer {\Rlam}
            {\Gamma,x{:}A; \cdot \entails e \against B} 
            {\Gamma; \cdot \entails \Lam{x}e \against A{\arr}B}
          ~~~
          \Infer {\Rapp}
            {\Gamma; \Delta_1 \entails e_1 \has A{\arr}B  ~&~  \Gamma; \Delta_2 \entails e_2 \against A}
            {\Gamma; \Delta_1, \Delta_2 \entails e_1 e_2 \has B}
        \end{tabular}
       }

        \\[1.4em]
        
         \begin{tabular}{r@{~~}l@{~~}l@{~~}l@{~~}l}
          \Infer {\Rsub}
            {\Gamma; \Delta \entails e \has A ~~&~~ A \subtype B}
            {\Gamma; \Delta \entails e \against B}
          ~~&~~
          \Infer {\Rfixvar}
            {\Gamma(u) = A}
            {\Gamma; \cdot \entails u \has A}
            &~~~
          \Infer {\Rfix}
            {\Gamma, u{:}A; \cdot \entails e \against A}
            {\Gamma; \cdot \entails \Fix{u}e \against A}
         \end{tabular}
        \\[1.4em]

         \begin{tabular}{rl}
           \Infer {\Rleftbot}
               {\Gamma \gokentails e \cok  ~~&~ \Delta, \X{:}\botty \linear e \cok}
               {\Gamma; \Delta, \X{:}\botty \entails e \against C}
            ~~&~~
           \Infer{\Rctxanno}
            {(\Gamma_0 \entails A) \ctxsubtype (\Gamma \entails A)
            ~~&~~  \Gamma; \Delta \entails e \against A}
            {\Gamma; \Delta \entails (e : \dots, (\Gamma_0 \entails A), \dots) \has A}
       \end{tabular}
        \\[1.4em]

          \begin{tabular}{r l}
           \fontsz{10pt}{\begin{tabular}{c}
              \Infer {\!\Rleftsect{1}}
                  {\Gamma; \Delta, \X{:}A \entails e \against C}
                  {\Gamma; \Delta, \X{:}A{\,\sectty\,}B \entails e \against C}
              \\[6pt]
              \Infer {\!\Rleftsect{2}}
                 {\Gamma; \Delta, \X{:}B \entails e \against C}
                 {\Gamma; \Delta, \X{:}A{\,\sectty\,}B \entails e \against C}
            \end{tabular}}\!\!
            &
            \!\!\! \vertcenter{
              \Infer{\Rsectintro}
               {\Gamma; \Delta \entails v \against A  ~~&~~  \Gamma; \Delta \entails v \against B}
               {\Gamma; \Delta \entails v \against A \sectty B}
           }
           \fontsz{10pt}{\begin{tabular}{c}
              \Infer {\!\Rsectelim{1}}
                {\Gamma; \Delta \entails e \has A \sectty B}
                {\Gamma; \Delta \entails e \has A}
              \\[6pt]
              \Infer {\!\Rsectelim{2}}
                {\Gamma; \Delta \entails e \has A \sectty B}
                {\Gamma; \Delta \entails e \has B}
            \end{tabular}}
       \end{tabular}
       \\~\\[-3pt]
       \begin{tabular}{rl}
           \Infer {\Rleftun}
             {\Gamma; \Delta, \X{:}A \entails e \against C
               ~~&~~ \Gamma; \Delta, \X{:}B \entails e \against C}
             {\Gamma; \Delta, \X{:}A \unty B \entails e \against C}
             &
            \Infer {\Runintro{1}}
                {\Gamma; \Delta \entails e \against A}
                {\Gamma; \Delta \entails e \against A \unty B}
           ~
            \Infer {\Runintro{2}}
                {\Gamma; \Delta \entails e \against B}
                {\Gamma; \Delta \entails e \against A \unty B}
       \end{tabular}
       ~\\[1.2em]
       $%
         \text{\begin{tabular}{c}
             \Infer{\Rdirectleft}
                {\arrayenvb{\Gamma; \Delta_1 \entails e' \has A}
               ~~&~  \Gamma; \Delta_2, \X{:}A \entails \E[\X] \against C}
               {\Gamma; \Delta_1, \Delta_2 \entails \E[e'] \against C}
               \textrm{\small where $e'$ is not a linear variable}
       \end{tabular}}%
       $
     \end{tabular}

     \vspace{-0.2em}
   \caption{The left tridirectional system}

   \label{fig:tri-system}
  \end{figure}

\subsection{Tridirectional typechecking and evaluation contexts}

Rule \Rdirectleft's use of an evaluation context might give the impression that
typechecking simply proceeds in the order in which terms are actually evaluated.  However,
this is not the case.  The subject of \Rdirectleft is $\E[e']$ where $e'$ synthesizes a
type, so certainly $e'$ must be in \emph{an} evaluation position, but there may be several
such positions.  Even a term as simple as $f~(x~y)$ has 5 subterms in evaluation position,
each corresponding to a different evaluation context $\E$:
\oneinferencerule{\begin{tabular}[t]{rcl}
    \begin{tabular}[t]{lcl}
        $\E = \hole~(x~y)$  & and & $e' = f$;
    \\[1pt]  $\E = f~(\hole~y)$  & and & $e' = x$;
    \end{tabular}
&
    \begin{tabular}[t]{lcl}
         $\E = f~(x~\hole)$  & and & $e' = y$;
     \end{tabular}
~~&~~
    \begin{tabular}[t]{lcl}
         $\E = f~\hole$  & and & $e' = (x~y)$;
    \\[1pt]  $\E = \hole$  & and & $e' = f~(x~y)$.
     \end{tabular}
\end{tabular}
}
In fact, we may need to repeatedly apply \Rdirectleft to the same subject term with
different choices of $\E$!  For example, we might use $\E = \hole~(x~y)$ to name an $f$ of
union type, introducing $\barvar{f}{:}A{\,\unty\,}B$ into the context; then, case-analyze
$A{\,\unty\,}B$ with \Runelim; finally, choose $\E = \barvar{f}~(\hole~y)$ to name $x$
(also of union type).  Thus we are faced not with a choice over decompositions, but over
many \emph{sequences} of decompositions.

Typechecking cannot go strictly left to right.  Given an ML-like \texttt{int\;option} type,
containing $\texttt{None}$ and some integer $\texttt{Some}(n)$, assume
$\texttt{None} : \None$ and $\texttt{Some}(n) : \SomeInt$.  Then, if
$\textfn{map}\;f\;(\texttt{Some}(n))$ returns $\texttt{Some}(f\;n)$
and $\textfn{map}\;f\;\texttt{None}$ is $\None$, then $\textfn{map} : 
(\Int \arr \Int) \arr ((\SomeInt \arr \SomeInt) \sectty (\None \arr \None))$.
Similarly, a function filtering out negative integers could have type $\textfn{filter} :
\Int \arr (\SomeInt \unty \None)$.

Consider
the term $
   (\textfn{map}~f)~(\textfn{filter}~n)
$.
\newcommand{\SSsNN}{\ensuremath{(\mathsf{s}{\arr}\mathsf{s}){\sectty}(\mathsf{n}{\arr}\mathsf{n})}}
The term $(\textfn{map}~f)$ synthesizes $(\SomeInt \arr \SomeInt){\sectty}(\None \arr \None)$.
This is an intersection type---we'll abbreviate it as \SSsNN---and the intersection
must be eliminated so that rule $\Rapp$ can be applied to $(\textfn{map}~f)~(\textfn{filter}~x)$.
However, we cannot commit to one part of the intersection yet, because we must first
case-analyze the union type of the subterm $(\textfn{filter}~x)$.  We need to ``jump over'' $(\textfn{map}~f)$
to type $(\textfn{filter}~x)$, so apply \Rdirectleft with evaluation context $\E =
\hole~(\textfn{filter}~x)$, giving $(\textfn{map}~f)$ the name $\X$; second, apply
\Rdirectleft with context $\E = \X~\hole$, synthesizing $\SomeInt \unty \None$ for
$(\textfn{filter}~x)$.  Rule \Rleftun splits on
$\Y{:}\SomeInt{\unty}\None$; in its left subderivation $\Dee_{\SomeInt}$, we have $\Y{:}\SomeInt$, so
\Rsectelim{1} on $\X \has (\SomeInt \arr \SomeInt) \sectty (\None \arr \None)$ gives
$\X \has \SomeInt \arr \SomeInt$, while its right subderivation $\Dee_{\None}$ has $\Y{:}\None$, so
\Rsectelim{2} gives $\X \has \None \arr \None$.  Writing $\Delta$ for $\X{:}(\SSsNN)$, the derivation is
\begin{displ}
   \Infer{\Rdirectleft}
      {\entails map~f \has \SSsNN
        &
       \Infer{\Rdirectleft}
           {\Delta \entails \textfn{filter}~x \has \SomeInt \unty \None
             ~&\hspace{-1pt}
              \Infer{\Rleftun} 
                   { \Dee_{\SomeInt}~&~
                       \Dee_{\None}
                    }
                   {
                     \Delta, \Y{:}(\SomeInt{\unty}\None) \entails \X~\Y \against C
                   }}
           {\Delta
                    \entails
                    \X~(\textfn{filter}~x) \against C
           }
       }
      {\entails (map~f)~(\textfn{filter}~x) \against C}
\end{displ}
where $C$ is $\SomeInt \unty \None$, and the derivations $\Dee_{\SomeInt}$ and $\Dee_{\None}$ are
\[
         \begin{array}[c]{c}\Infer{\Rsectelim{1}}
                                                  {\dots \entails \X \has (\SomeInt{\arr}\SomeInt){\sectty}(\None{\arr}\None)}
                                                  {\dots \entails \X \has \SomeInt \arr \SomeInt}
                                             \\[-7pt] \vdots \\[-4pt]
                                             \Delta, \Y{:}\SomeInt  \entails \X~\Y \against C
          \end{array}
      \ensuremath{~~\AND~~}
        \begin{array}[c]{c}\Infer{\Rsectelim{2}}
                                                  {\dots \entails \X \has (\SomeInt{\arr}\SomeInt){\sectty}(\None{\arr}\None)}
                                                  {\dots \entails \X \has \None \arr \None}
                                             \\[-7pt] \vdots \\[-4pt]
                                             \Delta, \Y{:}\None  \entails \X~\Y \against C
          \end{array}
\]
On a purely theoretical level, the tridirectional system is acceptable, but the
nondeterminism is excessive.  Xi approached (very nearly) the same problem by transforming
the program so the term of $\unty$ type appears before the term of $\sectty$ type.
(Actually, Xi had index-level quantifiers $\ExisSym$ and $\UnivSym$ instead
of $\unty$ and $\sectty$, but these are analogous.)
A standard let-normal translation $\xitrans{e}$~\cite[p.\ 86]{XiThesis}, where
$
     \xitrans{e_1\,e_2} = \Let{x_1}{\xitrans{e_1}}{\Let{x_2}{\xitrans{e_2}}{x_1x_2}}
$
suffices for the examples above.  (In Xi's system, existential variables are unpacked
where a term of existential type is let-bound: an existential variable $b'$ is unpacked at the binding of $x_2$,
which appears before the application $x_1x_2$ at which the universal variable $a$ must be instantiated.)
Unfortunately, the translation interacts unpleasantly with bidirectionality: terms such as
$\textfn{map}~(\Lam{x} e)$, in which $(\Lam{x} e)$ must be checked, no longer typecheck
because the $\lambda$ becomes the right hand side of a $\xxLet$, in
$
   \Let{x_1}{\textfn{map}}{
    \Let{x_2}{\Lam{x} e}{
     x_1 x_2}}
$ and let-bound expressions must synthesize a type, but $\Lam{x} e$ does not.
Typechecking becomes incomplete in the sense that some programs that were well typed
are not well typed after translation. %

Xi ameliorated this incompleteness by treating $e_1\,v_2$ as a special case~\cite[p.\ 139]{XiThesis}:
$ \xitrans{e_1\,v_2} = \Let{x_1}{\xitrans{e_1}}{x_1\,v_2}
$.
Now $v_2$ (which is $\Lam{x} e$ in the above example) is in a checking position.  This is
adequate for non-synthesizing values, but terms such as $\textfn{map}~(\Case{z}{\dots})$,
where a non-synthesizing non-value is in checking position, remain untypable.  It is not
clear why Xi did not also have special cases for $\keyword{case}$ and other
non-synthesizing non-values, \eg $\xitrans{e_1\,(\Case{e}{ms})} ~=~
\Let{x_1}{\xitrans{e_1}}{x_1\,\xitrans{\Case{e}{ms}}}$.
Xi's translation is also incomplete for terms like $f~(\Case{x}{ms})$.  Suppose
$x$ synthesizes a union that must be analyzed to select the appropriate part of
an intersection in the type of $f$.  Since $x$'s scope---and thus the scope of its union---is entirely
within the $\xxLet$ created for the $\xCase$, typechecking fails.
\begin{displ}
  \begin{array}[t]{rl}
  \xitrans{f\,(\Case{x}{ms})} &=~~ \Let{f_1}{f}{\Let{x_0}{\xitrans{\Case{x}{ms}}}{f_1~x_0}}
\\                                                 &=~~ \Let{f_1}{f}{\Let{x_0}{\big(\Let{x_1}{x}{\Case{x_1}{\xitrans{ms}}}\big)}{f_1~x_0}}
  \end{array}
\end{displ}

It could be argued that the cases in which Xi's translation fails are rare in practice.
However, that may only increase confusion when such a case is encountered.
I follow Xi's general approach of sequentializing the program before
typechecking, but no programs are lost in my translation.

Do we need all the freedom that \Rdirectleft provides?  No.  At the very least,
if we do not need to name a subterm, naming it anyway does no harm.  But
naming all the subterms only slightly reduces the nondeterminism.
Clearly, a strategy of in-order traversal is sound (we can choose to apply \Rdirectleft
from left to right if we like).  It is tempting to think it is complete.  In fact, it
holds for many programs, but fails for a certain class of annotated terms.  We will
explain why as we present the general mechanism for enforcing a strategy of left-to-right
traversal except for certain annotated terms.

\section{Let-Normal Typechecking}  \Label{sec:let}

We'll briefly mention previous work on let-normal form, then explain the ideas behind the
variant here, including why we need a \emph{principal synthesis of values} property.
Because the most universal form of principality does not hold for a few terms, we
introduce \emph{slack bindings}.

Traditional let-normal or A-normal
transformations~\citep{Moggi88:ComputationalLambda,Flanagan93:CompilingWithContinuations} 
(1) explicitly sequence the computation, and (2) name the result of each intermediate computation.
(Continuation-passing style
(CPS)~\citep{Reynolds93:DiscoveriesOfContinuations} also (3) introduces named continuations.
Thus let-normal form is also known as \emph{two-thirds CPS}.)
Many compilers for functional languages use some kind of let-normal form to facilitate optimizations;
see, for instance, \citet{Tarditi96:TIL}, %
\citet{Reppy01:LocalCPSConversion}, %
\citet{Chlipala05}, and
Peyton Jones \etal~(\citeyear{GHC:ANF}).

Our let-normal form will sequentialize the computation, but
it does not only name intermediate \emph{computations}, but \emph{values} as
well.  In our let-normal type system, \Rdirectleft is replaced by a rule \Rlet that can
\emph{only} be applied to $\xxLet$; see \Figureref{fig:let-system}.  $\Elong$ is a special evaluation context, discussed below.
   \begin{figure}[t]
   \centering
     \begin{tabular}{@{}l}
       \begin{tabular}[c]{@{}c@{}}
           $\Infer{\!\Rlet}
            {\Gamma; \Delta_1 \entails e' \has A
              ~&~ \Gamma; \Delta_2, \X{:}A \entails \Elong[\X] \against C}
            {\Gamma; \Delta_1, \Delta_2 \entails \Let{\X}{e'}{\Elong[\X]} \against C}$    
           \\[1em]
              \Infer{\!\Rletslack}
                {\Gamma; \Delta, \slackbind{\X}{v} \entails \Elong[\X] \against C}
                {\Gamma; \Delta \entails \Letslack{\X}{v}{\Elong[\X]} \against C}
            ~~~~~~
              \Infer{\!\Rslackvar}
                {\Gamma; \Delta_1 \entails v \has A
                  &~ \Gamma; \Delta_2, \X{:}A \entails e \against C}
                {\Gamma; \Delta_1, \Delta_2, \slackbind{\X}{v} \entails e \against C}
          \end{tabular}
   \\[-0.5em] ~\\
    \textit{\dots plus all rules in \Figureref{fig:tri-system}, except \Rdirectleft}
     \end{tabular}

     \vspace{-0.3em}
   \caption{The \emph{let-normal type system} for terms containing $\xxLet~\X$ bindings}

   \label{fig:let-system}
  \end{figure}

This \xxLet is a syntactic marker with no computational character.  In contrast to
let-normal translations for compilation purposes, there is no evaluation step (reduction)
corresponding to a $\xxLet$.  I won't even give a dynamic semantics for terms
with $\xxLet$s.  It would be easy; it's simply not useful here.  If we insist on knowing
what a let-normal term $e$ means, we can use a standard call-by-value operational semantics 
over the term's reverse translation. %

Instead of making explicit the order of computation, our let-normal form makes explicit
the order of \emph{typechecking}---or rather, the order in which \Rdirectleft names
subterms in evaluation position.  Thus, to be complete with respect to the tridirectional
system, the transformation must create a $\xxLet$ for every subterm in synthesizing form:
if an (untranslated) program contains a subterm $e'$ in synthesizing form, it might be
possible to name $e'$ with \Rdirectleft, so the let-normal translation must bind $e'$.
Otherwise, a chance to apply \Rleftun is lost. Even variables $x$ must
be named, since they synthesize a type and
so can be named in \Rdirectleft.   This models an ``aggressive'' strategy of
applying \Rdirectleft.  On the other hand, checked terms like $\Lam{x}e$ can't
synthesize, so we won't name them. 

Another consequence of the let-normal form following typing, not evaluation, is
that $\Let{\X}{v_1}{v_2}$ is considered a value---after all, the original term $[v_1/\X]\,v_2$
was a value, and we transformed a value into a non-value we could
not apply value-restricted typing rules such as \Rsectintro, leading to incompleteness.

We define the translation by a judgment $e \lettrans L + e'$, read
``$e$ translates to a sequence of let-bindings $L$ with body $e'$''.
For example, the translation of $f~(x~y)$, which names every synthesizing subterm,
is
\begin{displ}
   $~~~~\Let{\barvar{f}}{f}
     \Let{\X}{x}{
       \Let{\Y}{y}{
         \Let{\barvar{z}}{\X~\Y}{
           \Let{\barvar{a}}{\barvar{f}~\barvar{z}}{
             \barvar{a}
    }}}}$
\end{displ}
This is expressed by the judgment $
   f\;(x~y)   \,\lettrans\,
     \bind{\barvar{f}}{f},
     \bind{\X}{x},
     \bind{\Y}{y},
     \bind{\barvar{z}}{\X~\Y},
     \bind{\barvar{a}}{\barvar{f}~\barvar{z}}
     \,+\,
     \barvar{a}
$.
\Figureref{fig:lettrans} has the definition.
Note that $L + e'$ is not a term; $+$ is punctuation.
We write $L \freein e'$ as shorthand: read
$e \lettrans L + e'$ as ``$e \lettrans L \freein e'$''.  The divergent
notations come from the multiple decompositions of a term into
a pair of bindings and a ``body''.  For example,
$\Let{\barvar{x}_1}{e_1}{\Let{\barvar{x}_2}{e_2}{e_3}}$ can
be written three ways:
\textbf{(1)} $\cdot \freein \Let{\barvar{x}_1}{e_1}{\Let{\barvar{x}_2}{e_2}{e_3}}$, 
\textbf{(2)} $(\bind{\barvar{x}_1}{e_1}) \freein \Let{\barvar{x}_2}{e_2}{e_3}$, or
\textbf{(3)} $(\bind{\barvar{x}_1}{e_1}, \bind{\barvar{x}_2}{e_2}) \freein e_3$.
The last decomposition is \emph{maximal}: it has the maximum number of
bindings (and the smallest `body'), which is the case when the body isn't
a $\textkw{let}$.  If $e \lettrans L + e'$ then
$L \freein e'$ is maximal. %

Again, to model a complete strategy of \Rdirectleft-application,
in $e \lettrans L + e'$ we need $L$ to bind all the synthesizing subterms that could
be in evaluation position (after applying \Rdirectleft zero or more times).

We syntactically partition terms into \emph{pre-} and \emph{anti-}values.
A \emph{pre-value} $\PreV$ is a value, such as $x$, or a term that can ``become'' a value
via \Rdirectleft, such as $x~y$ which ``becomes'' the value $\Z$ in the derivation.  
(The h\'a\v{c}ek \raisebox{-3pt}{$\mathsz{16pt}{\check{~}}$} above the $e$ is shaped like a `v' for `value'.)
An  \emph{anti-value} $\AntiV$, such as $\Fix{u}{e}$ (or $\Case{e}{ms}$)
is not a value and cannot become a value.

\Rdirectleft can replace any synthesizing subterm with a linear variable, so
the pre-values must include both the values and the synthesizing forms.
This leads to the following grammar for pre-values, with values $x$, $\X$, and $\Lam{x} e$  %
and synthesizing forms $(e : As)$, $e_1 e_2$, $u$.  %
(In the full system, the prevalues also include checking forms that can become values if
all their subterms can, such as $(e_1, e_2)$.) %
\begin{displ}
  \begin{array}[t]{rcl}
  \text{Pre-values}~~~    \PreV ~&::= &~x ~|~ \X ~|~ (e : As) ~|~ \Lam{x} e ~|~ e_1 e_2  ~|~ u
\\\text{Anti-values}~~~      \AntiV ~&::= &~%
   \Fix{u}{e}
   \end{array}
\end{displ}
The distinction matters for terms with sequences of
immediate subterms such that at least two subterms in the sequence may be in evaluation position.
Only application $e_1 e_2$ has this property (and in the full system, pairs $(e_1, e_2)$).
$\Lam{x} e$ and $\Fix{u} e$ have no subterms in evaluation position at all.

A telling example is $(\Fix{u}{e})~(\omega~x)$ where $\omega : \dots
\arr \botty$.  In the tridirectional system, this term has no synthesizing subterms
in evaluation position.  In particular, $\omega~x$ is not in evaluation
position, so however we translate the term, we must not bind $\omega~x$ outside the outer application;
if we did, we would add $\Z{:}\botty$ to the context and could apply rule \Rleftbot to
declare the outer application well typed while ignoring $e$!  If $e$ is ill-typed, this is
actually unsound.  On the other hand, in the term $(f~g)~(\omega~x)$ the left
tridirectional system \emph{can} bind $\omega~x$ before checking the pair, by applying
\Rdirectleft with $\E = \hole\;(\omega~x)$ (synthesizing a type for $f~g$, ensuring soundness)
to yield a subject $\X\;(\omega~x)$ in which $\omega~x$ is in evaluation position.

The difference is that $\Fix{u}e$ is an anti-value, while $f~g$ is a
pre-value.  Therefore, given an application $e_1\;e_2$, if $e_1$ is some anti-value $\AntiV_1$,
the translation places the bindings for subterms of $e_2$ (\eg
$\bind{\Z}{\omega~x}$ above) \emph{inside} the argument part.  On the other hand, if
$e_1$ is a pre-value $\PreV_1$, the translation puts the bindings for subterms of $e_2$ \emph{outside} the application.
See the shaded rules in \Figureref{fig:lettrans}.

\emph{Elongated evaluation contexts} $\Elong$, unlike ordinary
evaluation contexts $\E$, can skip over pre-values.  $\Elong$ is a sort of
transitive closure of $\E$: if, by repeatedly replacing pre-values
in evaluation position with values, some subterm is then in evaluation position, that
subterm is in elongated evaluation position.  In a sequence of
\Rdirectleft-applications, subterms in evaluation position are replaced with linear
variables, which are values.  For example, $z$ is not in evaluation position in $(x~y)~z$,
but applying \Rdirectleft with $\E = \hole~z$ yields a subderivation with subject $\X~z$,
in which $z$ \emph{is} in evaluation position.  A $\Elong$ is thus a path that can skip
pre-values: if every intervening subterm is a pre-value (equivalently, if there is no
intervening anti-value), the hole is in elongated evaluation position.
The grammar for let-normal terms ensures that
the body $e_2$ of $\Let{\X}{e_1}{e_2}$ must have the form $\Elong[\X]$.

  \begin{displ}
    \begin{array}[t]{r r l l}
 \text{Elongated}&     \Elong &::=~~
         \hole ~|~ \Elong e ~|~ \PreV \Elong
~|~ (\Elong : As)
\\[2pt] \text{evaluation contexts} &&~~~~~~|~ \Let{\X}{\Elong}{e} ~|~ \Let{\X}{\PreV}{\Elong}
           ~|~ \Letslack{\X}{\Elong}{e} ~|~ \Letslack{\X}{v}{\Elong}
\\[4pt] \text{Terms} &     e &::=~~ \ldots
           ~|~ \Let{\X}{e_1}{\Elong[\X]}
           ~|~ \Letslack{\X}{v_1}{\Elong[\X]}
\\[3pt] \text{Values} & v &::=~~ x  %
           ~|~ \Lam{x}e %
           ~|~ \X
           ~|~ \Let{\X}{v_1}{v_2}
           ~|~ \Letslack{\X}{v_1}{v_2}
\\[3pt] \text{Eval. contexts}&   \E &::=~~ \ldots
           ~|~ \Let{\X}{\E}{e}
           ~|~ \Let{\X}{v}{\E}
           ~|~ \Letslack{\X}{\E}{e}
           ~|~ \Letslack{\X}{v}{\E}
~\\[3pt]
\text{Sequences of bindings} &     L &::=~~ \cdot
           ~|~ L, (\bind{\X}{e})
           ~|~ L, (\slackbind{\X}{v})
    \end{array}
  \end{displ}

\begin{figure}[t]

\begin{center}\!\!\!\!\!
   \begin{tabular}{c}
\multicolumn{1}{l}{\judgbox{ e \lettrans L + e'  }
     \raisebox{2pt}{read ``$e$ translates to bindings $L$ with result $e'$''}} \\[4pt]
    \InferAnon
           {}
           {x \lettrans (\bind{\X}{x}) + \X}
    ~~~~~~~
    \InferAnon
        {e \lettrans L+e'}
        {\Lam{x} e \lettrans \cdot + \Lam{x} (L \freein e')}
   \\[1em]
    \InferAnon
           {}
           {u \lettrans (\bind{\X}{u}) + \X}
   ~~~~~~~
    \InferAnon
        { e \lettrans L+e'}
        { \Fix{u} e \lettrans \cdot + \Fix{u} (L \freein e')}
    \\[1em]
    \shadebox{
        \InferAnon
           { \AntiV_1 \lettrans L_1+e_1'  ~~~&~~~    e_2 \lettrans L_2+e_2' }
           { \AntiV_1 e_2 \;\lettrans\; L_1, \bind{\X}{e_1' (L_2 \freein e_2')} + \X  }
    }
    ~~~~~
    \shadebox{
        \InferAnon
           { \PreV_1 \lettrans L_1+e_1'  ~~~&~~~    e_2 \lettrans L_2+e_2' }
           { \PreV_1 e_2 \;\lettrans\; L_1, L_2, \bind{\X}{e_1' e_2'} + \X  }
    }
   \\[1em]
    \InferAnon
       {    e \lettrans L+e'  ~~~&~~ e~\text{not a value}}
       {    (e : As) \lettrans L,\bind{\X}{(e' : As)} + \X}
   ~~~~~~~~~~
    \InferAnon
        {      v \lettrans L+e'}
        {      (v : As) \lettrans L, \slackbind{\X}{(e' : As)} + \X}
    ~~~~~~~~~~~
    \InferAnon
        {}
        {\X \lettrans \cdot + \X}
  \end{tabular}
\end{center}
~\\[-24pt] $~$
\caption{The let-normal transformation}
\label{fig:lettrans}
\end{figure}

\subsection{Principal synthesis of values} \Label{sec:let:principal-synthesis}

A key step in completeness is the movement of let-bindings
outward.   To prove this preserves typing, we show that principal
types~\citep{Hindley1969:Principal} exist in certain cases.  Consider the judgment
$
  x : (A_1{\arr}B) \sectty (A_2{\arr}B),
  y : A_1 \unty A_2; \cdot
  \entails 
  x~y  \against  B
$.
To derive this in the left tridirectional system, we need \Rdirectleft
with $\E = x~\hole$ to name $y$ as a new linear variable $\Y{:}A_1 \unty A_2$.  Then we
use \Rleftun; we must now derive
\begin{displ}
  $x : (A_1{\arr}B) \sectty (A_2{\arr}B),
  \dots; \Y{:}A_1
  \entails
  x~\Y \against B
~~~\AND~~~
  x : (A_1{\arr}B) \sectty (A_2{\arr}B),
  \dots; \Y{:}A_2
  \entails
  x~\Y \against B$
\end{displ}
Here, the scope of $\Y$ is $x~\Y$, and we synthesize a type for $x$ \emph{twice},
once in each branch:
\begin{displ}
\Infer{\Rdirectleft}{
  \dots, y : A_1 \unty A_2; \cdot \entails y \has A_1 \unty A_2
  &
  \Infer{\Rleftun}
    {\Infer{\Rapp}
        {\shadebox{\dots; \cdot \entails x \has A_1{\arr}B}
         &~~ \vdots~}
        {\dots; \Y{:}A_1 \entails x~\Y \against B}
      &
      \Infer{\Rapp}
         {\shadebox{\dots; \cdot \entails x \has A_2{\arr}B}
         &~~ \vdots~}
        {\dots; \Y{:}A_2 \entails x~\Y \against B}}
    {\dots; \Y : A_1 \unty A_2 \entails x~\Y \against B}
}{%
  x : (A_1{\arr}B) \sectty (A_2{\arr}B),
  y : A_1 \unty A_2; \cdot
  \entails 
  x~y  \against  B
}
\end{displ}
However, when checking the translated term
$
   \Let{\X}{x}{
   \Let{\Y}{y}{
   {\Let{\Z}{\X~\Y}{
   \Z
   }} }}
$
against $B$, we need to first name $x$ as $\X$, then $y$ as $\Y$,
then use \Rleftun to decompose the union $\Y{:}A_1 \unty A_2$ with
subject $\Let{\Z}{\X~\Y}{\Z}$.
\oneinferencerule{
\Infer{\Rlet}{
  \dots; \cdot \entails \shadebox{x \has (A_1{\arr}B) \sectty (A_2{\arr}B)}
  ~&~
\dots; \X{:}(A_1{\arr}B) \sectty (A_2{\arr}B) \entails \Let{\Y}{y}{
   {\Let{\Z}{\X~\Y}{
   \Z
   }} } \against B
}{%
  x : (A_1{\arr}B) \sectty (A_2{\arr}B),
  y : A_1 \unty A_2; \cdot
  \entails 
  \Let{\X}{x}{
   \Let{\Y}{y}{
   {\Let{\Z}{\X~\Y}{
   \Z
   }} }}  \against  B
}}
But we only get one chance (highlighted above) to synthesize a type for $x$,
so we must take care when using \Rlet to name $x$; if we choose to synthesize $x \has
A_1{\arr}B$ in \Rlet, we can't derive
\[
  \X{:}A_1{\arr}B, \Y{:}A_2 \entails (\Let{\Z}{\X~\Y}{\Z}) \against B
\]
but if we choose to synthesize $x \has A_2{\arr}B$ we
can't get 
\[
  \X{:}A_2{\arr}B, \Y{:}A_1 \entails (\Let{\Z}{\X~\Y}{\Z}) \against B
\]
  The only choice that works is $\Gamma(x)$, which is $(A_1{\arr}B) \sectty (A_2{\arr}B)$, since
given $\X \has (A_1{\arr}B) \sectty (A_2{\arr}B)$ we can synthesize $\X \has A_1 \arr B$ and
$\X \has A_2 \arr B$ using \Rsectelim{1} and \Rsectelim{2}, respectively.

In the above situation, $e' = x$ is a variable, so there is
a best type $C$---namely $\Gamma(x)$---such that if $x \has C_1$ and
$x \has C_2$ then $x \has C$, from which follows (by rules \Rsectelim{1,2} in the
example above)
$x \has C_1$ and $x \has C_2$.  We'll say that $x$ has the property of
\emph{principal synthesis}.  Which
terms have this property?  %
Variables do: %
the best type for some $x$ is $\Gamma(x)$.
On the other hand, it does not hold for many non-values: $f~x \has A_1$ and $f~x \has A_2$
do \emph{not} imply $f~x \has A_1 \sectty A_2$, since the intersection introduction rule
\Rsectintro is (1) restricted to values and (2) in the checking direction.  Fortunately,
we don't need it for non-values: Consider $(e_1\:e_2)~y$.  Since
$(e_1\:e_2)$ is not a value, $y$ is not in evaluation position in $(e_1\:e_2)~y$, so even
in the tridirectional system, to name $y$ we must first name $(e_1\:e_2)$.  Here, the
let-normal system is no more restrictive.  Moreover, some
values, such as pairs, are checking forms and \emph{never} synthesize, so they do not have
the principal synthesis property.  But neither system binds values in checking form to linear variables.

Now, do all \emph{values in synthesizing form} have the principal synthesis property?
The only values in synthesizing form are ordinary variables $x$, linear variables
$\X$, and annotated values $(v : As)$.  For $x$ or $\X$ the principal type is simply
$\Gamma(x)$ or $\Delta(\X)$.  Unfortunately, principal types do not always exist for
terms of the form $(v : As)$.  For example,
$  %
    ((\Lam{x} x) : (\entails \Unit \arr \Unit), (\entails \Tbool \arr \Tbool))
$ %
can synthesize $\Unit \arr \Unit$, and it can synthesize $\Tbool \arr \Tbool$, but
it can't synthesize their intersection, so it has no principal type.

\subsection{Slack bindings}

Rather than restrict the form of annotations, we use a different
kind of binding for $(v : As)$---a \emph{slack binding}
$\slackbind{\X}{v}$ where $v$'s type is synthesized not at its binding site,
but at any point up to its use (rules \Rslackvar and \Rletslack in \Figureref{fig:let-system}).
Wherever $\X$ is in scope, we can try rule \Rslackvar to synthesize a
type $A$ for $v$ and replace $\slackbind{\X}{v}$ with an ordinary linear variable
typing $\X{:}A$.  For example, $\big((\Lam{x} e) : (\entails \Int{\arr}\Int)\big)~y$ is translated to
$
\Letslack{\X}{\big((\Lam{x}e') : (\entails \Int{\arr}\Int)\big)} \Let{\Y}{y} \X\;\Y
$.  
Now, we have several chances to use \Rslackvar to synthesize the type of $\X$: just before
checking $\Let{\Y}{y} \X\;\Y$, or when checking $\X\;\Y$.
This is just like choosing when to apply \Rdirectleft in the tridirectional system.
If all our bindings were slack we would have put ourselves in motion to no purpose,
but we'll use slack bindings for $(v : As)$ \emph{only}.  My experiments
suggest that slack bindings are rare in practice~\citep[p.\ 187]{DunfieldThesis},
and are certainly less problematic than the backtracking from intersections and
unions themselves (\Rsectelim{1,2}, etc.).
%
%
%
%

\iffalse
    \subsubsection{Remark on $\topty$}
      The above discussion of best typings centred on $\sectty$, but a similar issue arises
      with $\topty$.  Suppose $e_2 \has \botty$.  The let-normal translation of $((\Lam{x} x) : {\,}\entails \Unit)~e_2$ is
      \[
        \Let{\X}{((\Lam{x} x) : {\,}\entails \Unit)}{
          \Let{\Y}{e_2}{
           \Let{\Z}{\X~\Y}{
             \Z
       }}}
      \]
      The original term is well typed in the tridirectional system:
      $e_2 \has \botty$ is in evaluation position, so we can apply \Rdirectleft and then use
      \Rleftbot to show that the entire term checks against any type whatsoever (this is
      perfectly sound: if $e_2 \has \botty$ then $e_2$ diverges, so $v~e_2$ diverges for all
      $v$, including $((\Lam{x} x) : \dots)$).  But in a let-normal system without slack
      variables, we are stuck because $((\Lam{x} x) : {\,}\entails \Unit)$ does not synthesize
      anything.  A solution might be to behave as though $\topty$ were always one of the
      annotations, so $(v : As) \has \topty$ for all $v$, but slack variables work as well:
    $
        \Letslack{\X}{((\Lam{x} x) : {\,}\entails \Unit)}{
          \Let{\Y}{e_2}{
           \Let{\Z}{\X~\Y}{
             \Z
       }}}
    $.
      Here the offending term $((\Lam{x} x) : {\,}\entails \Unit)$ is bound to $\X$ but rule
      \Rslackvar is never applied, since typechecking is ``short-circuited'' by \Rleftbot.
\fi

%
%

%
\section{Results} \label{sec:results}

The two major results are \emph{soundness}: if the let-normal translation of a program is
well typed in the let-normal type system, the original program is well typed in the left
tridirectional system---and \emph{completeness}: if a program is well typed in the left
tridirectional type system, its translation is well typed in the let-normal type system.
Once these are shown, it follows from \citet{Dunfield04:Tridirectional} that the
let-normal system is sound and complete with respect to a system~\citep{Dunfield03:IntersectionsUnionsCBV}
for which preservation and progress hold under a call-by-value semantics.

At its heart, the let-normal system merely enforces a particular pattern of linear variable introductions
(via \Rlet, instead of \Rdirectleft).  So it is no surprise that soundness holds.
The proof is syntactic, but not too involved; see \citet[pp.\ 132--134]{DunfieldThesis}.

\begin{corollary*}[Let-Normal Soundness] %
  ~\\
  If $e \lettrans L + e'$ and $\cdot; \cdot \entails L \freein e' \against C$
  (let-normal system) then $\cdot; \cdot \entails e \against C$ (tridirectional system).
\end{corollary*}
\ifproofs
\begin{proof}
  By \Lemmaref{lem:lettrans-let-respecting}, $L \freein e'$ is let-respecting.
  By \Theoremref{thm:let-soundness},
  $\cdot; \cdot \entails \unw{L \freein e'} \againstL C$.  Substituting $e$ for
  $\unw{L \freein e'}$, justified by \Propositionref{prop:unwinding-is-inverse},
  yields the result.
\end{proof}
\fi

However, completeness---that the let-normal system is not
\emph{strictly} weaker than the tridirectional system---is involved.
What follows is the roughest sketch of the proof found in \citet[pp.\ 135--165]{DunfieldThesis}.
We want to show that given a well-typed term $e$, the let-normal translation
$L \freein e'$ where $e \lettrans L
+ e'$ is well-typed.  To be precise, given a derivation
$\Dee$ deriving $\Gamma; \Delta \entails e \against C$ in the left tridirectional system,
we must construct a derivation
$\Gamma; \Delta \entails L \freein e' \against C$ in the let-normal system, where
$e \lettrans L + e'$.  My attempts
to prove this by straightforward induction on the derivation failed:
thanks to \Rdirectleft, the relationship between $e$ and $\Dee$
is complex.  Nor is $L + e'$ compositional in $e$: for a given
subterm of $e$ there may not be a corresponding subterm of $L + e'$, because translation
can insert bindings inside the translated subterm.

Instead, the completeness proof proceeds as follows:

\begin{enumerate}
\item \emph{Mark} $e$ with \keyword{let}s wherever \Rdirectleft is used in $\Dee$.
  However, if \Rsectintro or another subject-duplicating rule is used, the subderivations need not
  apply \Rdirectleft in the same way, resulting in distinct terms to which \Rsectintro
  cannot be applied.  So we use step 2 inductively
  to obtain typing derivations for the canonical version of the subterm (the $L + e'$ from
  $e \lettrans L + e'$), to which
  \Rsectintro can be applied.  

  This step centres on a lemma which produces a
  term with a let-system typing derivation.  This term might not be canonical.
  For example, if the original tridirectional derivation for
  $\Lam{x} x$ didn't use \Rdirectleft at all, no \keyword{let} bindings are created,
  unlike the canonical let-normal term $\Lam{x} \Let{\X}{x}{\X}$.
  
\item \emph{Transform} the marked term into the canonical $L + e'$  in small steps,
  adding or moving one \textkw{let} at a
  time.  Each small step preserves typing.  We'll define a syntactic measure $\mu$ that quantifies how
  different a term is from $L + e'$; each \textkw{let}-manipulating step reduces the measure, bringing the term closer to $L +
  e'$.  When the measure is all zeroes, the
  term \emph{is} $L + e'$. %
\end{enumerate}

  The \emph{measure} of $e'$ is
   \[
     \mu(e') ~=~ \langle
                            \mathsf{unbound_{\has}}(e'),~
                            \mathsf{brittle}(e'),~
                            \mathsf{prickly}(e'),~
                            \mathsf{transposed}(e')
                      \rangle
  \]
  where:
  \begin{itemize}
  \item   $\mathsf{unbound_{\has}}(e')$ is the number of subterms of $e'$
      in synthesizing form (that is, variables $x$ and $u$, annotated terms $(e : As)$, and applications $e_1 e_2$)
      that are not let-bound.  The translation $\lettrans$ has let-bindings for all such terms, so an $e'$ that does not bind such terms
      is quite far from being in canonical let-normal form.
  \item  $\mathsf{brittle}(e')$ is the number of let-bindings in $e'$ of the form $\Let{\X}{(v_1 : As)}{e_2}$.  To
    correspond to the translation $\lettrans$, we need to change such let-bindings to slack bindings $\Letslack{\X}{(v_1 : As)}{e_2}$.  These terms are ``brittle'' because they need to be slackened.
  \item  $\mathsf{prickly}(e')$ is the number of let-bindings in $e'$ that are not properly collected together at a \emph{root}.
    A \emph{root} is somewhere that the canonical translation $\lettrans$ may place a sequence
    of let-bindings.  In the proof, we start by reducing the number of unbound synthesizing forms by inserting $\xxLet$s nearby,
    but some of these are too deep inside the term.  For example, given a term $c~y$, we first put a binding around the
    $y$, giving $c\,(\Let{\Y}{y}{\Y})$.  (To simplify the example, $c$ is some constant or primitive operation that
    is never let-bound.)  Then we bind the application, giving $\Let{\barvar{a}}{c~(\Let{\Y}{y}{\Y})}{\barvar{a}}$.
    But the canonical translation would be $\Let{\Y}{y}{\Let{\barvar{a}}{c~\Y}{\barvar{a}}}$.  Thus, a prickly binding
    needs to be lifted outward until it is in some sequence of let-bindings at the outside of the body of a $\xLam$ or $\xFix$,
    or at the outside of the entire term $e'$.
  \item  $\mathsf{transposed}(e')$ is the number of \emph{transposed variable pairs} in $e'$.
    If there are no prickly bindings, there may still be bindings that are out of order.  For a term $x~y$, the original
    derivation might have used \Rdirectleft first on $y$ (with $\E = x~\hole$) then on $x$ (with $\E = \hole~\Y$).
    In this case, Step 1 above would produce $\Let{\Y}{y}{\Let{\X}{x}{\X~\Y}}$.  Supposing this application is the body
    of some $\xLam$, these bindings are not prickly, but don't correspond to what $\lettrans$ would produce.  Variables
    (and their bindings) are transposed if they are not used in the same order they were bound.  Thus, $\X$ and $\Y$ are
    transposed in $\Let{\Y}{y}{\Let{\X}{x}{\X~\Y}}$, because $\Y$ is bound before $\X$ but $\X$ appears to the left of $\Y$
    in the body of the $\xxLet$.
  \end{itemize}

We interpret the quadruples lexicographically.  Likewise, the proof of completeness relies on type preservation lemmas
for each part of the quadruple: adding a let-binding preserves typing, changing a regular let-binding to a slack let-binding
preserves typing, lifting a let-binding to a root preserves typing, and reordering the bindings of transposed variables
preserves typing.

\begin{theorem*}[Let-Normal Completeness]  %
  ~\\  If $\cdot; \cdot \entails e \against C$ (tridirectional system) and $e \lettrans L + e^*$
  then $\cdot; \cdot \entails L \freein e^* \against C$ (let-normal system).
\end{theorem*}

\section{Related Work}

The effects of transformation to continuation passing style on the precision of program
analyses such as 0-CFA have
been studied for some time~\citep{Sabry94:CPS}.  The effect depends on the specific details
of the CPS transform and the analysis
done~\citep{Damian01:SyntacticAccidents,Palsberg03:CPS}. 
The ``analysis'' in this work is the process of bidirectional checking/synthesis.  My
soundness and completeness results show that my let-normal transformation does not
affect the analysis.  It is not clear if this means anything for more traditional
let-normal transformations and compiler analyses.

\section{Conclusion}

Transforming programs into a let-normal form removes a major impediment to implementing
tridirectional typechecking.  The system is sound \emph{and complete} with respect to a
type assignment system for intersections and
unions~\citep{Dunfield03:IntersectionsUnionsCBV}, in contrast to systems~\citep{XiThesis}
in which completeness is lost.  The tridirectional rule \emph{can} be turned into something practical.  A
chain of soundness results~\citep[p.\ 165]{DunfieldThesis} guarantees that if we run a
program $e$ whose let-normal translation typechecks in the system in this paper, it will
not go wrong.

Despite ``untangling'' \Rdirectleft, typechecking is still very time-consuming in the
worst cases, thanks to checking terms several times in \Rsectintro and backtracking in
\Rsectelim{1,2}, etc.  As implementing (an extended version of) this system
shows~\citep{Dunfield07:Stardust}, bad cases do occur in practice!

Parametric polymorphism is absent, but I have extended the tridirectional system and the
let-normal implementation~\citep{Dunfield09:polymorphism}, and the
soundness and completeness results should still hold.

The major flaw of this work is its completeness proof, which uses purely syntactic methods,
is complicated, and has not been mechanized.  Ideally, it would be mechanized and/or
proved more simply.

\paragraph*{Acknowledgments}
{\raggedright Many thanks to Frank Pfenning for countless discussions
about this research.  \\ Thanks also to the ITRS reviewers.
Most of the work was done at Carnegie Mellon University with
the support of the US National Science Foundation.}

\bibliography{letpaper}
\bibliographystyle{plainnat}
\end{document}
